\documentclass[a4paper,fleqn]{cas-dc}

\usepackage[numbers]{natbib}
\usepackage{amsmath}
\usepackage[ruled,vlined]{algorithm2e}
\usepackage[capitalise]{cleveref}
\usepackage{booktabs}
\usepackage{pifont}
\usepackage{adjustbox}
\usepackage{subcaption}
\usepackage{graphicx}

\def\tsc#1{\csdef{#1}{\textsc{\lowercase{#1}}\xspace}}
\tsc{WGM}
\tsc{QE}
\tsc{EP}
\tsc{PMS}
\tsc{BEC}
\tsc{DE}

\begin{document}
\let\WriteBookmarks\relax
\def\floatpagepagefraction{1}
\def\textpagefraction{.001}

\shorttitle{MARTINI}

\shortauthors{Nweye and Nagy}

\title [mode = title]{MARTINI: Smart Meter Driven Estimation of HVAC Schedules and Energy Savings Based on WiFi Sensing and Clustering}

\author[1]{Kingsley Nweye}[style=english,orcid=0000-0003-1239-5540]
\ead{nweye@utexas.edu}
\credit{Conceptualization, Data curation, Formal analysis, Methodology, Visualization, Roles/Writing - original draft, Writing - review \& editing}

\author[1]{Zoltan Nagy}[style=english,orcid=0000-0002-6014-3228]
\ead{nagy@utexas.edu}
\cormark[1]
\cortext[cor1]{Corresponding author}
\credit{Conceptualization, Formal analysis, Investigation, Methodology, Project administration, Resources, Supervision, Validation, Visualization, Roles/Writing - original draft, Writing - review \& editing}

\affiliation[1]{organization={Intelligent Environments Laboratory\\Department of Civil, Architectural and Environmental Engineering\\
  The University of Texas at Austin},
    addressline={301 E. Dean Keeton St., ECJ 4.200}, 
    city={Austin},
    postcode={78712-1700}, 
    state={Texas},
    country={USA}
}

\begin{abstract}
HVAC systems account for a significant portion of building energy use. Nighttime setback scheduling is an energy conservation measure where cooling and heating setpoints are increased and decreased respectively during unoccupied periods with the goal of obtaining energy savings. However, knowledge of a building's real occupancy is required to maximize the success of this measure. In addition, there is the need for a scalable way to estimate energy savings potential from energy conservation measures that is not limited by building specific parameters and experimental or simulation modeling investments. Here, we propose MARTINI, a sMARt meTer drIveN estImation of occupant-derived HVAC schedules and energy savings that leverages the ubiquity of energy smart meters and WiFi infrastructure in commercial buildings. We estimate the schedules by clustering WiFi-derived occupancy profiles and, energy savings by shifting ramp-up and setback times observed in typical/measured load profiles obtained by clustering smart meter energy profiles. Our case-study results with five buildings over seven months show an average of 8.1\%--10.8\% (summer) and 0.2\%--5.9\% (fall) chilled water energy savings when HVAC system operation is aligned with occupancy. We validate our method with results from building energy performance simulation (BEPS) and find that estimated average savings of MARTINI are within 0.9\%--2.4\% of the BEPS predictions. In the absence of occupancy information, we can still estimate potential savings from increasing ramp-up time and decreasing setback start time. In 51 academic buildings, we find savings potentials between 1\%--5\%.
\end{abstract}

\begin{highlights}
\item MARTINI derives ramp-up and ramp-down times of HVAC systems based on measured occupancy schedules
\item Savings estimates are provided, even if no occupancy data is available
\item Daily and seasonal occupancy and chilled water profile identified respectively.
\item 8.1\%--10.8\% (summer) and 0.2\%--5.9\% (fall) chilled water average energy savings for UT campus
\item Saving estimates are within 0.9--2.4\% projections of building energy performance simulation.
\end{highlights}


\maketitle
\section{Introduction}
Residential and commercial sectors are responsible for 40\% of energy consumption in the United States~\cite{eia-mer}, while Heating, Ventilation and Air Conditioning (HVAC) alone account for 44\% of energy use in commercial buildings~\cite{eia-cbecs}. These highlight the opportunities for energy conservation measures (ECMs) to improve HVAC system energy efficiency in commercial buildings. Commonly applied ECMs are high-performance building envelopes, lighting and HVAC controls \cite{Qian2019}. Nighttime setback (and morning ramp-up) is an HVAC ECM where the air handling units (AHUs) in the building are operated in energy savings mode, e.g., higher/lower temperature set-points for cooling/heating, minimum ventilation rates, etc, during unoccupied periods. For optimal scheduling of the AHUs, i.e., maximize energy savings while maintaining comfortable conditions for occupants, knowledge of the building's real occupancy is required.

Occupancy information is typically based on the facility manager's experiential knowledge of a building or on standardized occupancy schedules determined by the American Society for Heating, Refrigeration and Air Conditioning Engineers (ASHRAE) \cite{ashrae_90_1}. These approaches fail to take into account the stochastic nature of occupants or the fact that no two buildings function in the same way. The ubiquity of Wireless Fidelity (WiFi) infrastructure and connected devices provide cost-effective and non-intrusive means to collect building occupancy information, which can be used to develop building-specific occupancy schedules~\cite{Labeodan2015}. There is also, the need to estimate energy savings rewards from ECMs before or after implementation. Field experiments and building energy performance simulation are the typical go-to approach for such estimation purposes but possess significant investments for actualization. The deployment of smart meters in commercial buildings has given rise to a large influx of building energy data at high temporal resolution, which have been used for forecasting and fault detection~\cite{miller2015}. However, widespread utilization of such algorithms is challenging in practice due to the often involved complexity and data management.

The objective of our work is to develop a methodology that proposes an ECM and estimates its potential energy savings solely using smart meter and WiFi data. We introduce MARTINI: a sMARt meTer drIveN estImation of HVAC schedules and energy savings. In brief, we estimate a) typical occupancy schedules based on clustering of WiFi-derived occupancy information, and b) energy savings based on a virtual demand profile consisting of shifting setback and ramp-up times observed in typical/measured load profiles obtained by clustering smart meter data. By aligning setback and ramp-up times with actually observed occupancy, we can minimize energy use without impacting comfort. The main advantage of MARTINI is that it leverages readily available data and can be applied easily to a large number of buildings.  

This paper is organized as follows; \cref{sec:literature_review} reviews the related work, while \cref{sec:methodology} describes MARTINI. Then, in \cref{sec:case_study}, we apply MARTINI to five university buildings, and estimate their energy savings potential. We compare MARTINI's savings estimations to that of a detailed building energy model, and extrapolate savings to an additional 46 buildings where occupancy information is not available. We discuss our results in \cref{sec:discussion} followed by our concluding statements in \cref{sec:conclusion}. 
\section{Literature Review} \label{sec:literature_review}
\subsection{Occupant Information Sensing}
Spatio-temporal occupant information properties are popularly used in building HVAC control and consist of presence, count, location, track and identity \cite{thiago2010}. Presence is a binary property describing if there is at least one occupant present in the sensed environment. Count property seeks to determine the number of present occupants in the environment and is relevant to demand-driven cooling, heating and ventilation control. Location of an occupant is easily inferred when presence is known but the resolution of localization is limited by the coarseness of the sensing infrastructure for example, occupancy sensors located in every thermal zone versus one occupancy sensor at the main entry point of a building. Identity and track properties seek to answer the questions of \textit{What is the person doing?}, \textit{Who is each person?} and \textit{Where was this person before?} respectively \cite{thiago2010} and are most suitable for personalized occupant-centric control (OCC).

Various occupancy information sensing systems have been used for HVAC systems control including CO\textsubscript{2}, Passive Infrared (PIR), image (camera) sensors and, electromagnetic signals (WiFi, Bluetooth). CO\textsubscript{2} and PIR sensors are readily available in commercial buildings and have been used extensively for occupant presence and count sensing \cite{GRUBER2014548,JIANG2016132,FRANCO2020101714,5708933,NEWSHAM2017137}. However, CO\textsubscript{2} sensors have been found to have delayed response to concentration changes and are sensitive to sensor location, occupant activity and density, location of windows and doors \cite{ALISHAHI2021107936}. On the other hand, the use of PIR sensors is dependent on occupant motion and not necessarily presence resulting in false negatives when occupants are present but stationary e.g. sitting at a workstation \cite{PARK2019397}. Vision-based sensing, using cameras has been explored by researchers as a reliable alternative to CO\textsubscript{2} and PIR sensors \cite{WANG2017155,Liu_2013} however, privacy concerns remain an issue. This may be tackled through lower resolution imaging that preserves occupants' identity \cite{9314180}. Asides deployment costs, it is also computationally expensive to use cameras as they require intensive image processing. Electromagnetic signals in the form of Bluetooth or WiFi signals have given birth to various frameworks for occupancy information detection \cite{Park2018ABB,ZOU2017633,LONGO2019106876,8049459,cisco_meraki_2021}. This sensing system is advantageous because it uses non-intrusive methods that leverage existing wireless network infrastructure such as WiFi routers and mobile devices to passively infer occupant information.

WiFi derived occupant counts have been shown to have strong correlation with actual ground truth counts where R\textsubscript{2} values ranging from 0.85 - 0.96 have been reported in the literature \cite{hobson2020,simma2019,8255034,8049459} as well as a strong occupant to device ratio of 1.27 \cite{8843224}. WiFi-based occupancy can be utilized in solving occupancy estimation, future occupancy prediction and occupancy pattern clustering problems \cite{ALISHAHI2021107936}. In solving the occupancy estimation problem, \cite{ZOU2017633} achieved 98.85\% occupancy detection accuracy, 0.096 Normalized Root Mean Square Derivative occupancy count accuracy and 1.385 m occupancy tracking accuracy using their framework that utilized Received Signal Strength (RSS) levels and device MAC addresses. \cite{8255034} achieved 90\% accuracy using Channel State Information (CSI) to infer occupant counts where the sensitivity of transmitted radio signals to occupant movements was captured. Their approach eliminated the need for occupant terminal devices such as smartphones. \cite{OufMohamedM2017EouW} compared the estimated occupancy using CO\textsubscript{2} sensorS to WiFi networks and found that WiFi networks provided higher accuracy and had lower implementation cost. \cite{9312032} explored supervised classification and regression models in predicting both occupant presence and counts respectively in an academic building and achieved up to 86.69\% accuracy and 0.29 Root Mean Square Percentage Error. The authors in \cite{8843224} sought to develop regression models for 24 hours in advance occupant count prediction using Multi-Layer Receptor (MLR) and Artificial Neural Network (ANN) supervised regression models and achieved 83.1\% and 90.1\% accuracy on the test sets of the MLR and ANN models respectively. 

\subsection{WiFi-sensed Occupant Information for HVAC Scheduling}
A campus-scale occupancy prediction framework for HVAC scheduling was proposed by \citeauthor{trivedi2017} where they trained WiFi-derived occupancy counts from 112 campus buildings on an ensemble gradient boosting regressor and achieved  95.35\% model accuracy. Their large dataset captured a stronger spatio-temporal variation in occupancy compared to other studies that investigated fewer buildings \cite{9312032,8843224}. However, they only reported the missed and waste time from using static schedules over their proposed WiFi-derived schedules but did not provide an estimate on actual energy savings. Also, the adoption of their framework in an actual building may be challenging because of BAS system memory storage limitations and the associated risks of granting external access to a BAS network. Other researchers have sought to identify characteristic occupancy patterns in buildings using unsupervised machine learning techniques which are better than typical static schedules and easier to implement in practice compared to supervised regression models. Using KMeans clustering, \cite{hobson2020} found five representative clusters from the WiFi occupancy data of an academic office building, which where then used to develop a decision tree classification model for day-ahead occupancy forecasting. The occupancy forecasts were utilized in an HVAC scheduling application. Compared to \cite{trivedi2017} their model is easier to program in a BAS since it can be simplified to if-else conditional statements. Also, a decision tree is easier to communicate to building stakeholders from different fields. Similarly, our work employs KMeans clustering in estimating representative occupancy schedules to be used in nighttime setback scheduling, and we seek to generate actionable results that can be implemented readily.

\subsection{Energy Savings Estimation Methods}
There is the need to quantify the energy savings from HVAC ECMs for communication with building stakeholders prior to or after implementation. These estimates are arrived at through experimentation, building energy performance simulation (BEPS) or data-driven methods. \cite{BalajiBharathan2013SobH} controlled the temperature setback in 55 of 237 HVAC zones in their test building using WiFi derived occupancy information, and measured 17.8\% and  2.2\% electrical and cooling energy savings respectively from one day of experimentation. However, they were constrained by the number of occupants able to participate in the experiment and the baseline and experimental conditions were not perfectly aligned in terms of weather and occupancy. In their review, \cite{ParkJuneYoung2019Acro} cited communication challenges between researchers and facility managers, as well as the need to recommission equipment as some of the challenges of OCC field experimentation. Moreover, there could be security and comfort concerns associated with manipulating HVAC systems during field experimentation that also limit the duration these experiments are able to last for, resulting in outcomes that are difficult to generalize.

\begin{figure*}
  \centering
  \includegraphics[width=\textwidth]{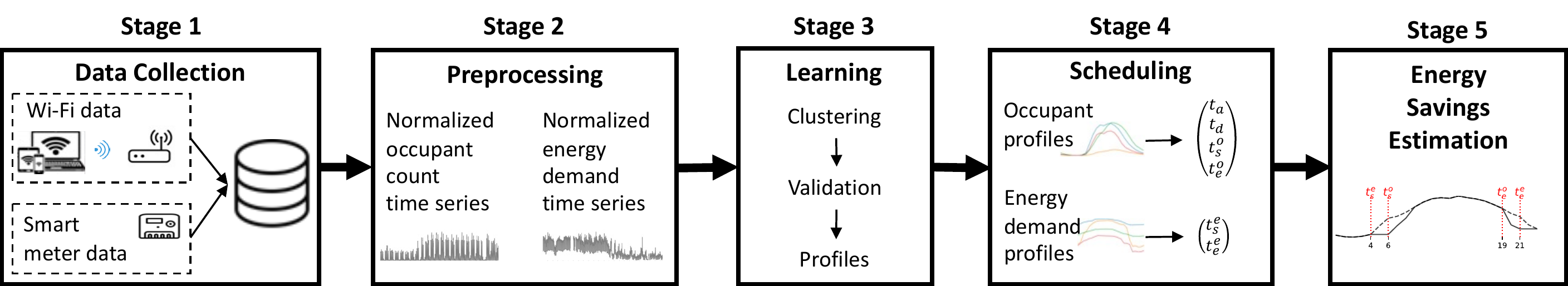}
  \caption{MARTINI overview.}
  \label{fig:methodology-overview}
\end{figure*}

Building energy performance simulation (BEPS) addresses some of the challenges of field experimentation as it does not require direct occupant participation, allows for multiple simulations with different design scenarios under similar climatic conditions, and is a deterministic modeling approach able to capture the governing physics of a building's energy use. BEPS has been used to estimate energy savings from OCC-focused ECM applications \cite{AZAR2012841,marschall2019,CHONG2021116492} using whole-building simulation software such as EnergyPlus \cite{crawley2000energy} and eQUEST \cite{hirsch2006equest}. \cite{simma2019b} applied three different occupancy schedules in a lecture hall's energy model with the first being a fixed schedule, the second being a ground truth occupancy-based schedule and the third being a WiFi-derived occupancy schedule. The simulations showed 57\% and 56\% energy savings using the ground truth occupancy-based and WiFi-derived schedules over the fixed schedule. A major challenge in BEPS is the model development cost involving collation of building envelope and systems parameters that may not be readily available. To tackle this, simulations are done on a smaller scale by either focusing on a single building or few zones. Other software like Windows Air Model (WinAM) aimed at Continuous Commissioning\copyright{ }of existing buildings \cite{esl_cc}, reduce the modeling complexity by allowing the designer to focus primarily on the air-side and plant-side building definitions. A major simplification feature and somewhat compromise of WinAM is that it does not account for solar gains and thermal mass in its algorithm and it calculates energy use one hour at a time, independent of the previous hour. This allows for faster computation and fewer user inputs \cite{likins2018comparison}. WinAM has been used to evaluate ECM applications in building HVAC systems and showed outcomes that are comparable to those gotten from other modeling software, e.g., EnergyPlus ~\cite{yang2013study,likins2018comparison}. Here, we use WinAM to validate the energy savings proposed by MARTINI.

The use of data driven techniques for building energy prediction is well known and widely used \cite{FOUCQUIER2013272,deb2017review,amasyali2018review,zhao2012review}. The surge in smart energy monitoring devices has increased the amount of available building energy data, making data-driven methods possible for building energy performance analysis \cite{grillone2020}. A deterministic modeling approach like BEPS is building specific, which makes it difficult to cost-effectively make estimations on a large scale \cite{GEYER201732}. Data-driven methods have the potential to be applied easier across many buildings. In their review on ECM energy savings estimation methods, \cite{grillone2020} listed user-facing fallen rule list using audit data, artificial neural networks using audit data, clustering techniques, linear regression and genetic algorithms trained on databases as some of the data-driven methods that have been used for prediction and ECM recommendations. However, \emph{no work exists that is capable of making ECM intervention propositions, alongside with estimated energy savings, based solely on WiFi and smart meter data}. This is the research gap that MARTINI is filling as we detail in \cref{sec:methodology}.
\section{Methodology: MARTINI} \label{sec:methodology}
An overview of MARTINI for nighttime setback scheduling and energy savings estimation is shown in \cref{fig:methodology-overview}, consisting of five stages. In stage 1, WiFi connection logs between WiFi-enabled devices and wireless access points (WAPs) are collected in all buildings and sent to a central database. Energy demand, metered at the building level using smart meters, is also sent to the central database. In stage 2, preprocessing steps such as data cleaning, transformation and aggregation are applied to the WiFi and smart meter data to produce building-level occupant count and energy demand time series. All data are then normalized using min-max normalization as
\begin{equation}
  T_{norm}(i) = \frac{T(i) - \min(T)}{\max(T) - \min(T)} 
  \label{eqn:min-max-normalization}
\end{equation}
where $T_{norm}(i)$ is the mapping of the value at index $i$ in vector $T$ to a value between 0 and 1, $\min(T)$ and $\max(T)$ are the respective minimum and maximum values in $T$.

In stage 3, we use KMeans to cluster daily occupancy and energy demand profiles. The optimal number of cluster partitions is determined using the elbow-method on the Within-Cluster-Sum of Squared Errors (WSS) 
\begin{equation}
  \textrm{WSS} = \sum^k_{i=1} \sum_{x \in C_i} (x - c_i)^2
  \label{eqn:wss}
\end{equation}
where $k$ is the number of clusters and $C_i$ are data points in the $i^\textrm{th}$ cluster with centroid $c_i$.

\begin{figure}
  \centering
  \includegraphics[width=.8\columnwidth]{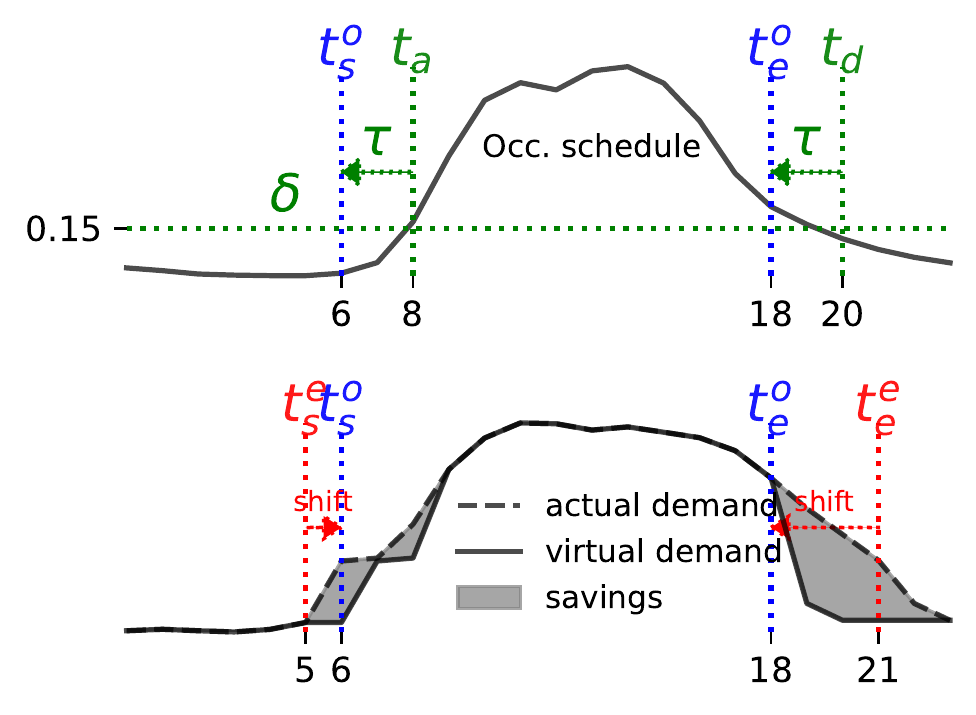}
  \caption{MARTINI example.}
  \label{fig:methodology_example}
\end{figure}

In stage 4, the occupancy and energy demand centroids from stage 3 are used as representative building occupancy schedules and energy demand profiles, respectively, to determine potential savings as follows (see \cref{fig:methodology_example}). We define the occupied threshold, $\delta$, as the minimum occupant density, i.e., normalized occupant count for an occupied state, and use it to extract first arrival time, $t_a(\delta)$, and last departure time, $t_d(\delta)$, from the occupancy schedules such that $t_a$ is the time when the occupancy density first rises above $\delta$ and $t_d(\delta)>t_a(\delta)$ is the time when the occupancy density drops below $\delta$. To account for a time lag $\tau$ due to thermal inertia of the building \cite{Pang2017}, we adjust $t_a$ and $t_d$ by $\tau$ to derive the occupant-derived HVAC ramp-up, $t_s^o$, and setback time $t_e^o$ as (see \cref{fig:methodology_example} top)
\begin{align}
    t_s^o(\delta,\tau) &= t_a(\delta) - \tau \\
    t_e^o(\delta,\tau) &= t_d(\delta) + \tau
\end{align}
Our central hypothesis is that if the building is operated within $t_s^o$ and $t_e^o$, unnecessary energy use is avoided without impacting occupant comfort. Notice that $\tau$ is a property of the building which can generally not be altered, $\delta$ is a choice of the operator to balance savings and comfort, and the $t_s^o$ and $t_e^o$ may vary for each cluster. In our experiments below, we use $\tau=2$hr and vary $\delta$.

Also in stage 4, we estimate the energy demand profile-derived ramp-up, $t_s^e$ and setback time, $t_e^e$, by calculating the difference between the energy demand profile vector and its negative LAG 1 (L\textsuperscript{-1}) vector so that the times having the largest negative and positive difference are $t_s^e$ and $t_e^e$ respectively. In other words, $t_s^e$ and $t_e^e$ represent the smart meter driven estimate of the building's operational (static) schedule, evaluated for each cluster on its centroid.

In the last stage, we estimate the energy savings as follows. First, we denormalize the data to get the demand in energy units. We then create a virtual demand profile for each day by translating the actual demand profile from $t_s^e$ to $t_s^o(\delta,\tau)$ (from the respective cluster of the day), assuming a constant demand on $[t_s^e, t_s^o(\delta,\tau)]$, and similarly for $t_e^o(\delta,\tau)$ and $t_e^e$. Integrating the difference between the actual and the virtual demand yields the daily energy savings. Repeating this procedure for every day yields the total energy savings for a building.

\cref{fig:methodology_example} shows an example application of MARTINI, for $\delta=0.15$ and $\tau=2\textrm{hr}$. From the WiFi occupancy counts (\cref{fig:methodology_example} top), we find $t_a(0.15)=\textrm{08:00}$ and $t_d(0.15)=\textrm{20:00}$. Thus, $t_s^o(0.15,2)=\textrm{06:00}$ and $t_e^o(0.15,2)=\textrm{18:00}$. Next, we find $t_s^e=\textrm{05:00}$ and $t_e^e=\textrm{21:00}$ from the smart meter data (\cref{fig:methodology_example} bottom). The virtual profile is created by translating the actual profile from $t_s^e$ to $t_s^o(0.15,2)$ and from $t_e^e$ to $t_e^o(0.15,2)$, and the resulting energy savings are shaded in grey.

It is pertinent to note that aligning energy demand with occupancy schedule may not always result in energy savings as it is possible to have occupied periods that are missed when unaligned, especially in the evenings. However, the net difference in energy demand from applying our method is what we term energy savings.
\section{Case-Study} \label{sec:case_study}
We present a real-world case-study of MARTINI where we use data from the University of Texas at Austin (UT) campus buildings located in Austin, Texas and classified under ASHRAE climate zone 2A. In total, we select 51 buildings composed of 6, 14 and 31 office, college laboratory, and college classroom, respectively. The buildings' build year and gross floor area are between 1894 - 2012, and 139m\textsuperscript{2} - 3,490m\textsuperscript{2} respectively. The period of analysis is 2019-07-09 to 2019-12-19, i.e., one summer and one fall semester. Of the 51 buildings, five have WiFi connection data and are used to estimate savings from aligning $t_s^e$ and $t_e^e$ with $t_s^o(\delta,\tau)$ and $t_e^o(\delta,\tau)$. We validate the savings estimations using WinAM building energy performance simulation software. Then, all 51 buildings are used to estimate the sensitivity of energy consumption with respect to shifted $t_s^e$ and $t_e^e$. This is the use-case for when occupancy information is not available, yet savings can be estimated. \cref{subsec:case_study-data_collection,subsec:case_study-preprocessing,subsec:case_study-learning,subsec:case_study-scheduling,subsec:case_study-energy_savings_estimation} detail the application of each of the stages 1—5 of MARTINI to our case study data.

\subsection{Data Collection} \label{subsec:case_study-data_collection}
\subsubsection{WiFi Connection Data}
WiFi enabled devices connect to Wireless Access Points (WAPs) and timeout every five minutes, and the connection data is uploaded to a centralized campus database. We retrieve the data, made up timestamp, building name, wireless access point name and hashed WiFi device Media Access Control (MAC) address fields, and store in our local database. The MAC address serves as a unique identifier for WiFi devices and is encrypted before being sent out to preserve occupant identity and maintain privacy. During this work's development, a new data stream was created that contained additional data fields including unique encrypted user identifiers which will be useful in solving the problem of count overestimation. This new data stream was however not used in this work because its earliest timestamp was 2019-12-17 which is near the end of the Fall 2019 semester and subsequent data after the 2019 Winter Break contain days that fall within the COVID-19 pandemic and do not depict typical occupancy patterns during normal operation. A total of 1,208 unique WAPs and 176336 hashed device MACs from five buildings (out of the 51) were available and collected.

\subsubsection{Smart Meter Data}
On the district cooling and heating network of the university, chilled water, steam and electricity use are metered using smart meters that provide five-minute resolution data at the building level. The university's Utility and Energy Management (UEM) provides a data pipeline to store this data in our local database. The raw data contains the timestamp and fields for each building's chilled water, steam and electricity demand. We only investigate chilled water energy savings in our analysis as our case study climate is cooling dominant.

Other data not explicitly relevant to MARTINI but used in BEPS validation of our method include building envelope and mechanical system data for the selected buildings and local weather data. Exterior wall R-value (thermal resistance), window U-value (thermal conductance), number of AHUs and types, operation schedules, design flow rates and temperature set points are acquired through a survey sent out to UEM staff where they filled out values for the aforementioned parameters based on their expert building and operational knowledge. We use floor plans to estimate building geometry while Google Earth is utilized for building elevation and fenestration area estimation. Camp Mabry (KATT) Austin, Texas weather station weather file was used for BEPS.

\subsection{Preprocessing} \label{subsec:case_study-preprocessing}
\subsubsection{WiFi Connection Data}
WAPs allow connection from all WiFi enabled devices including smartphones, laptops and wearable technology that may travel with the occupant in the building and desktop computers and Internet of Things (IOTs) which may be stationary and associated with not one occupant in particular. Also, devices belonging to passer-bys within or outside the buildings where the WAPs are housed can be connected. There is the unique situation where a WAP may be located on the exterior of a building or have no clear indication of its location in the set of available buildings. An occupant could also own multiple devices associated with different WAPs across a building. WAP names are formatted using a regular expression that contains the name of the building, floor and nearest room where the WAP is located. By finding instances of WAP names not matching the regular expression, externally located WAPs are identified. Externally located WAPs are not utilized in developing occupant counts. We identify three types of WiFi devices in a building as a function of their total daily connection minutes, $t$, as \cite{Park2018ABB,zhWang2019}.

Short-stay devices are attributed to passers-by and briefly visiting occupants having less than 45 minutes of WAP connection per day where 45 minutes is the minimum duration for a class lecture at the case-study university. Regular devices are attributed to the typical occupants of a building with 45 to 540 minutes of WAP connection per day where 540 minutes is the duration that would have been spent for a typical 08:00 - 17:00 work schedule and any device with connection minutes exceeding 540 minutes is assumed to be stationary i.e. non-mobile devices assumed not to travel along with occupants. This classification is based on knowledge of the buildings' space use being office and academic hence for other use types, new devices classes may be defined.

Device counts are then derived by aggregating the number of each unique device type in a building per timestamp. This results in five-minute resolution of building level occupant counts. The data counts are re-sampled to an hourly resolution with the counts rounded up to the nearest whole number and timestamps without data are assigned zero count. Other spatial-temporal occupant count resolution are possible to extract from the connection data as it is collected at 5-minute resolution and, the WAP IDs contain information about a WAP's floor and nearest room location. These may be used to estimate room and floor level occupancy counts for zone-level control. However, such fine spatial granularity suffers from uncertainty like changes in WAP location without updating to its name leading to wrong localization of an occupant. Also differing densities of WAPs across a building could cause overestimated counts at certain areas of a building.

\begin{figure}
  \centering
  \includegraphics[width=\columnwidth]{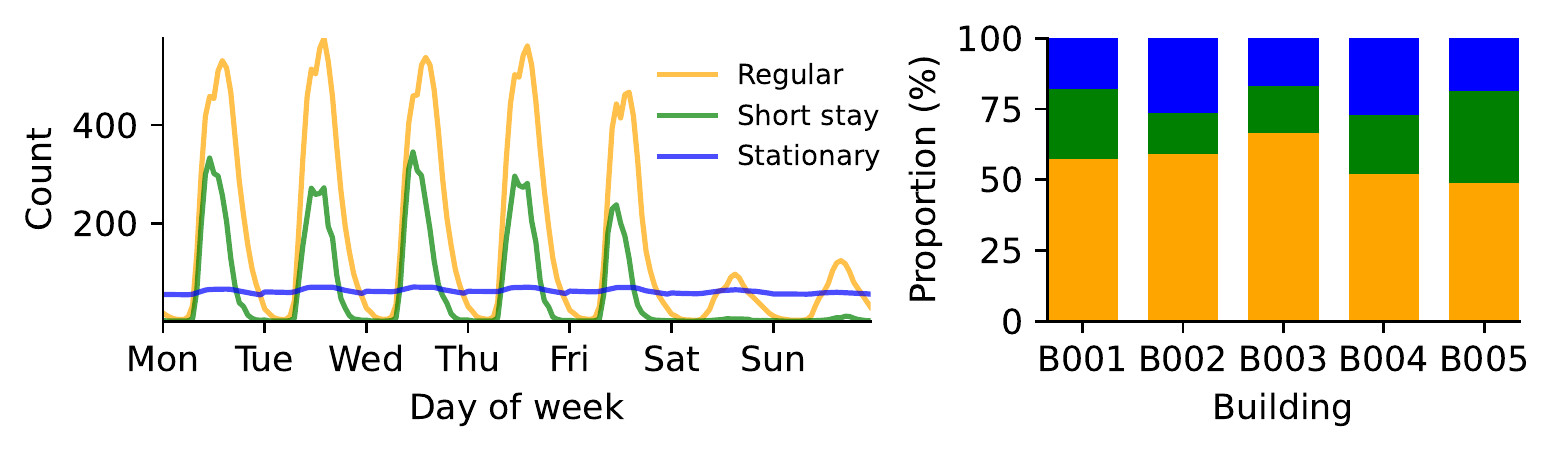}
  \caption{Device type distribution in WiFi connection data.}
  \label{fig:device_classification}
\end{figure}

\cref{fig:device_classification} shows the average weekly device type counts and their proportion per building in the WiFi connection data. Although regular and short-stay devices arrive at about the same time each day of the week, they peak at different times and regular devices leave later in the day. Also, there are almost no short-stay devices during weekends. As expected, stationary device counts remains nearly constant on an average week. The largest proportion of devices are regular at about 50\% or more in the five buildings. We use only the regular devices to produce the normalized occupant count time series.

\subsubsection{Smart Meter Data}
We down-sample the chilled water energy demand data from five-minute to hourly resolution to match that of occupant counts. Outliers exist in the energy demand data and include isolated spikes occurring in specific buildings or happening at the campus level such that it occurs in multiple buildings at the same time. Our discussions with UEM reveal that such anomalies are results of consumption values during periods of power outages, delay in data logging at the smart meters or delay in transmission to the central server that eventually upload as a single lump sum. Another observed outlier type described as a \textit{flat-line}, is constant logging of a low value, sometimes as low as 0, that could last for hours. This trend is usually preceded or followed by a spike in demand. It is very unlikely that a commercial building operates without cooling demand as a result of interior zones. To identify and remove the isolated spikes, we use the inter-quartile range (IQR) method for outlier detection. In each building, the inter-quartile range of daily demand peaks is used to define the validity boundary for the hourly demand values. All values greater than the IQR upper bound are nullified. \textit{flat-line} outliers are detected by identifying instances where the rolling sum of change in demand is constant and are nullified. Missing values in the energy demand data are then approximated using linear interpolation for instances that have valid values in preceding and following timestamps.

\subsubsection{Exploratory Data Analysis}
\begin{figure}
  \centering
  \includegraphics[width=\columnwidth]{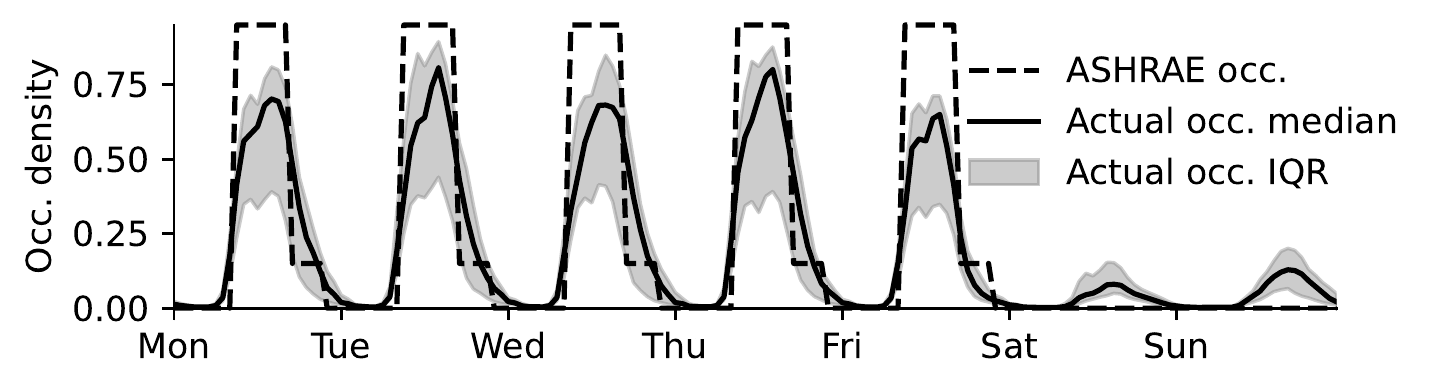}
  \caption{Comparison between WiFi-derived and ASHRAE school/university \cite{ashrae_90_1} weekly occupancy schedules for the five buildings having WiFi connection data.}
  \label{fig:wifi_occupancy_density_distribution_vs_ashrae_university}
\end{figure}

\begin{figure}
  \centering
  \includegraphics[width=\columnwidth]{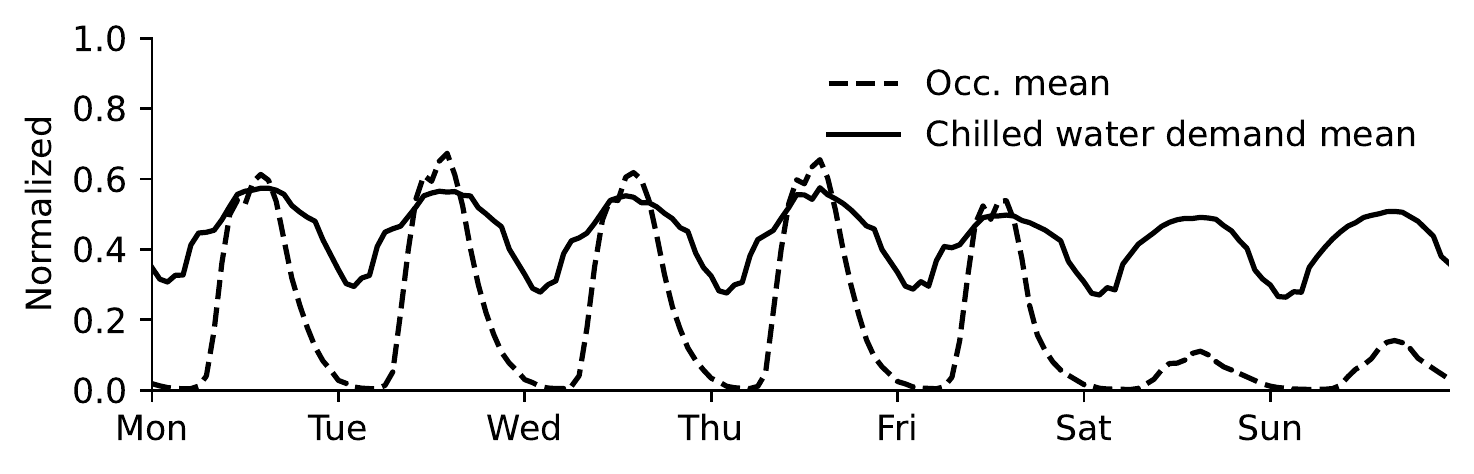}
  \caption{Comparison between normalized WiFi-derived occupancy schedule and chilled water demand for the five buildings having WiFi connection data.}
  \label{fig:norm_occupancy_vs_chilled_water}
\end{figure}

\cref{fig:wifi_occupancy_density_distribution_vs_ashrae_university} shows the difference between an average week's occupancy schedule derived from our WiFi data and ASHRAE's school/university building occupancy schedule \cite{ashrae_90_1}. The ASHRAE peak occupancy is higher, reached at an earlier time and at a faster rate each weekday compared to the WiFi-derived schedule. The ASHRAE schedule does not capture the temporal variation in occupancy as every weekday has the same schedule and buildings are assumed to be unoccupied on weekends. Also, the WiFi-derived schedule shows occupants arriving earlier and departing later than prescribed by ASHRAE. We highlight the disparity between average weekly chilled water energy demand profile and occupancy schedule from the five WiFi data buildings in \cref{fig:norm_occupancy_vs_chilled_water}. While the demand and occupancy peak almost simultaneously, the demand ramp-up and ramp-down occur much earlier and later respectively compared to when occupants arrive and depart. Although weekend occupancy is much lower than weekdays, the weekend chilled water energy demand magnitude is similar to that of weekdays. These observations motivate our initial hypothesis that WiFi derived occupancy schedules can lead to reduced energy use by designing for and operating within the actual occupancy schedule.

\subsection{Learning} \label{subsec:case_study-learning}
\subsubsection{Clustering}
The clustered data sets have shapes of $n \times 24$ where $n$ is the total number of daily profiles from all buildings and each data point has 24 variables describing the hourly occupant density or normalized chilled water demand. \cref{tab:data_summary} summarizes the case-study data sets with respect to their semester, number of buildings and data points i.e. daily profiles contained in data set. Occupancy (WiFi-derived) data are split into summer and fall semester sets denoted by IDs D-OS and D-OF respectively. Only five buildings are contained in each of these data sets. The chilled water demand data are also split into summer and fall semester sets denoted by IDs D-CS and D-CF and contain 51 buildings per set, including the five in D-OS and D-OF. The split occurs on the date of 2019-08-23 which is the beginning of UT's 2019 fall semester. Each of the data sets in \cref{tab:data_summary} is clustered independently using KMeans algorithm with the number of clusters ranging from $k=2\dots10$.

\begin{table}
    \centering
    \begin{tabular}{lllrr}
        \hline
        ID & Data set & Semester & \# Bldgs & \# Datapts \\ 
        \hline
        D-OS & Occ. & Summer & 5 & 225 \\
        D-OF & Occ. & Fall & 5 & 590 \\
        D-CS & CHW & Summer & 51 & 2121 \\
        D-CF & CHW & Fall & 51 & 5961 \\
        \hline
    \end{tabular}
    \caption{Case-study data summary.}
    \label{tab:data_summary}
\end{table}

\subsubsection{Validation}
\begin{figure}
  \centering
  \includegraphics[width=\columnwidth]{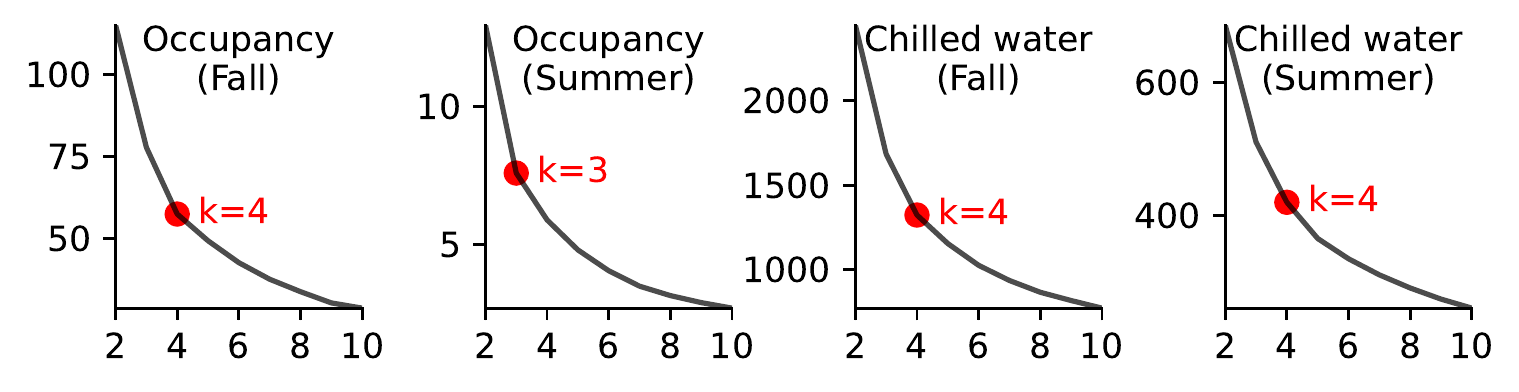}
  \caption{Within-Cluster-Sum of Squared Errors (WSS) for different $k$ number of cluster partitions, $k$ and selected optimal $k$ using elbow method.}
  \label{fig:cohesion_score}
\end{figure}

The WSS errors for $k=2\dots10$ clusters is shown in \cref{fig:cohesion_score}. Using the elbow method, $k=4$, is chosen to be optimal for D-OF, D-CS and D-CF data sets while $k=3$ is chosen to be optimal for D-OS data set as these $k$ values show the beginning of diminishing return on WSS.

\begin{figure*}
    \begin{subfigure}[]{.5\textwidth}
        \centering
        \includegraphics[width=\textwidth]{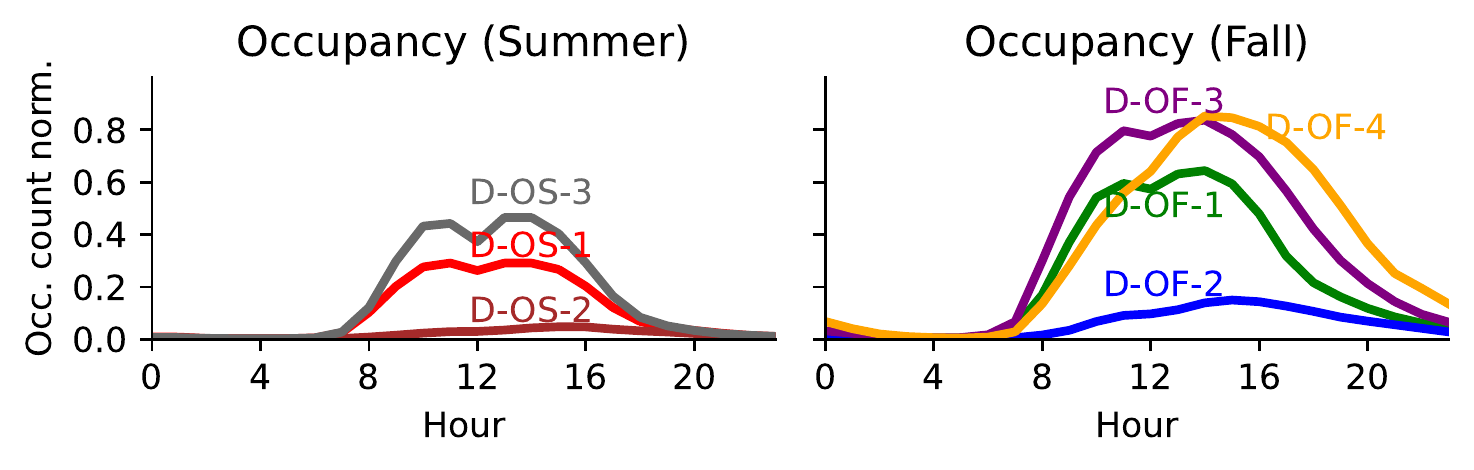}
    \end{subfigure}\hfill
    \begin{subfigure}[]{.5\textwidth}
        \centering
        \includegraphics[width=\textwidth]{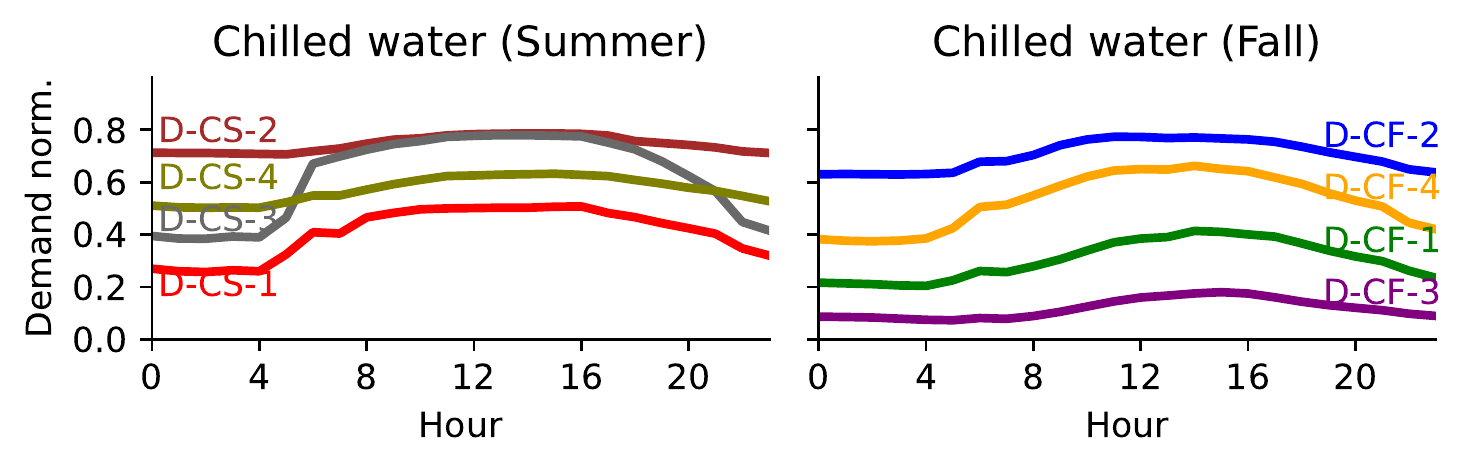}
    \end{subfigure}
    \caption{Cluster centroids used as representative occupancy schedule and chilled water demand profiles.}
    \label{fig:cluster_centroids}
\end{figure*}

\subsubsection{Profiles}
We utilize the cluster centroids as representative occupancy schedules and chilled water demand profiles. \cref{fig:cluster_centroids} shows the profiles corresponding to the optimal $k$ number of clusters for data sets D-OS, D-OF, D-CS and D-CF. Profile D-OS-2 and D-OF-2 can be described as low or unoccupied days where the normalized occupancy count is well below 20\%. D-OS-3 has a pronounced drop in occupancy by noon which may be attributed to occupants departing the buildings at lunchtime. D-OS-1 shows a slower occupancy build up but fewer departures at noon compared to D-OS-3. All three D-OS profiles have near-zero occupancy before 07:00 and after 21:00 and show early start of departure before 15:00. Compared to the D-OS profiles, D-OF profiles have higher occupant count peaks and show residual occupancy late at night and into the early hours of the day which could be attributed to more after-work activities in the buildings during fall semester. Profiles D-OF-1 and D-OF-3 have similar shapes compared to D-OF-4 which, has slower morning rise in occupancy and later start of departure. Nevertheless, D-OF-1, D-OF-3 and D-OF-4 have the same peak time of about 15:00. Most opportunities for energy savings come from summer days because of their early start of departure compared to fall days, and also fall days associated with profile D-OF-2 as the profile has the second lowest peak amongst D-OS and D-OF profiles. From the fall and summer chilled water profiles, D-CS-1 and D-CS-4 are similar in shape as the normalized demand does not show much variance. Profiles D-CS-1 and D-CS-3 show more pronounced ramp-up by their distinct rise in demand between 04:00--08:00. The fall chilled water profiles have very similar shapes but differ in peak magnitude. However, the start of ramp down is not clear from the chilled water profiles with the exception of D-CS-3 as a result of the gradual demand descent after midday.

\begin{figure}
    \centering
    \includegraphics[width=\columnwidth]{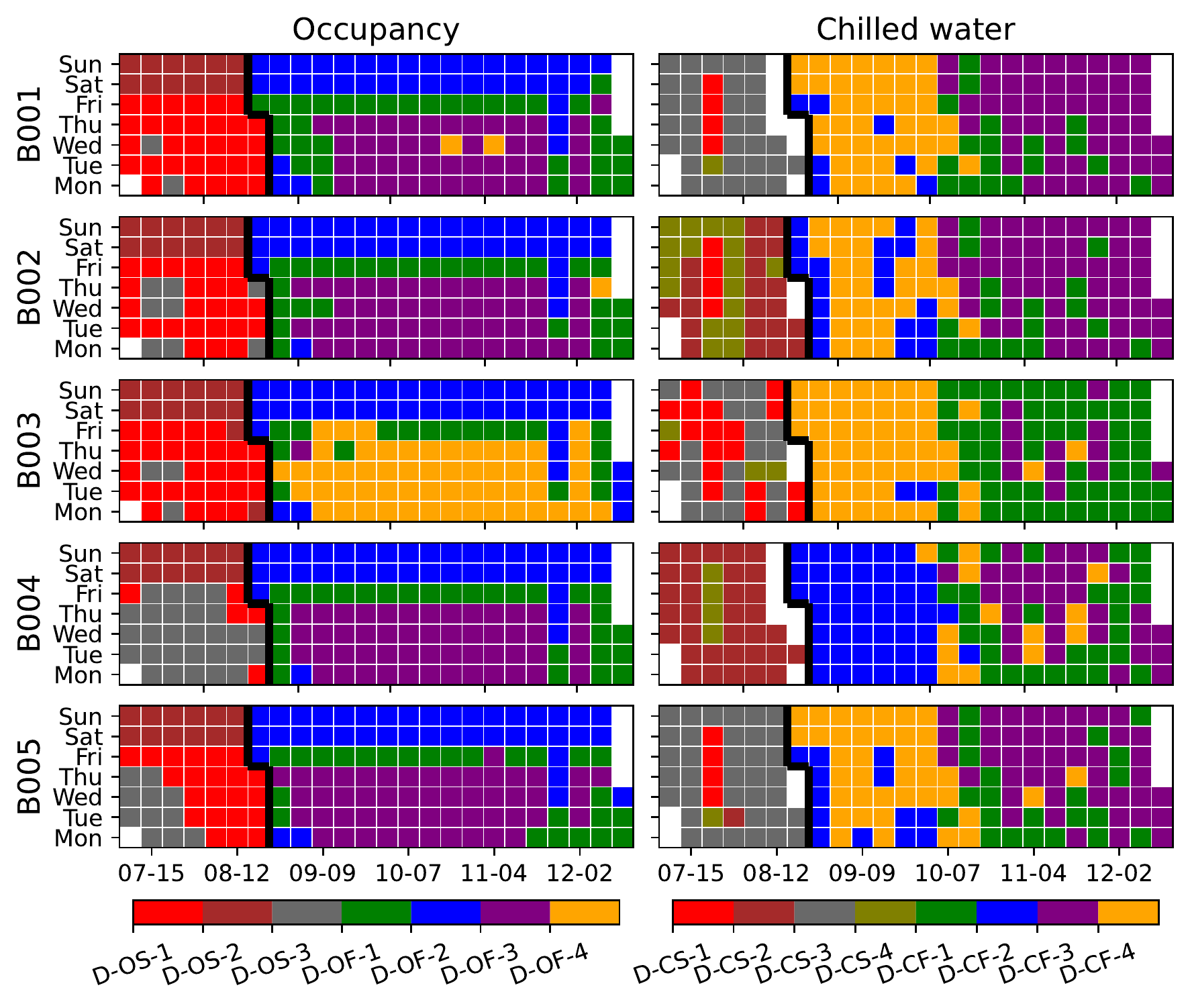}
    \caption{Daily occupancy schedule and chilled water demand profile cluster assignment. Black line demarcates summer from fall.}
    \label{fig:cluster_calendar}
\end{figure}

\cref{fig:cluster_calendar} shows the daily occupancy schedule and chilled water demand profile cluster assignment for the five buildings in the D-OS and D-OF data sets. We observe a weekly occupancy pattern in all buildings as there are distinct schedules for weekdays and weekends. All summer and fall weekend schedules are defined by D-OS-2 and D-OF-2 profiles respectively. Summer weekday occupancy schedules are majorly represented by D-OS-1 profile in buildings B001, B002 and B003 while B004 has a majority of its weekday occupancy profiles represented by D-OS-3 profile. Monday -- Thursday in B005 are described by D-OS-3 profile during the first two weeks before gradually transitioning to the D-OS-1 profile in subsequent weeks. The fall weekdays have distinct Friday occupancy schedule compared to Monday -- Thursday occupancy schedule. All five buildings show the same daily cluster pattern in the fall with the exception of building B003 whose Monday -- Thursday occupancy schedule is primarily defined by D-OF-3. These observations show that the clustering is robust in identifying unique building occupancy schedules. We outline each building's occupancy schedules in \cref{tab:building_occupancy_schedules} by taking the mode profile for Mondays -- Thursdays, Fridays and Saturdays -- Sundays per semester such that each building's occupancy schedule for the entire analysis period is defined by six profiles.

We observe seasonal pattern in chilled water demand as seen in \cref{fig:cluster_calendar}. For the time period of our analysis, chilled water had three distinct demand seasons; one in the summer and two during the fall. The major difference between the seasonal demand profiles is their magnitude where there is high demand in summer that diminishes towards the end of fall. We also observe unique building patterns for example, the demand profile in B004 during the beginning of fall is predominantly D-CF-2 whereas for other buildings, it is predominantly D-CF-4. We use the representative chilled water demand profiles in place of actual daily profiles for subsequent energy savings estimation.

\subsection{Scheduling} \label{subsec:case_study-scheduling}
\begin{table}
\centering
\caption{Building occupancy schedules.}
\label{tab:building_occupancy_schedules}
\begin{tabular}{lllll}
\toprule
Building &   Semester & Mon-Thu & Fri & Sat-Sun \\
\midrule
       B001 &  Summer &  D-OS-1 &  D-OS-1 &  D-OS-2 \\
      B002 &  Summer &  D-OS-1 &  D-OS-1 &  D-OS-2 \\
      B003 &  Summer &  D-OS-1 &  D-OS-1 &  D-OS-2 \\
       B004 &  Summer &  D-OS-3 &  D-OS-3 &  D-OS-2 \\
      B005 &  Summer &  D-OS-1 &  D-OS-1 &  D-OS-2 \\
       B001 &    Fall &  D-OF-3 &  D-OF-1 &  D-OF-2 \\
      B002 &    Fall &  D-OF-3 &  D-OF-1 &  D-OF-2 \\
       B003 &    Fall &  D-OF-4 &  D-OF-1 &  D-OF-2 \\
      B004 &    Fall &  D-OF-3 &  D-OF-1 &  D-OF-2 \\
      B005 &    Fall &  D-OF-3 &  D-OF-1 &  D-OF-2 \\
\bottomrule
\end{tabular}
\end{table}

\cref{tab:occ_time_signals} shows $t_a(\delta)$, $t_d(\delta)$, $t_s^o(\delta,\tau)$ and $t_e^o(\delta,\tau)$ inferred from the occupancy schedule profiles using our approach described in \cref{sec:methodology} where $\tau=2$ and, $\delta=0.05$, $\delta=0.1$ and $\delta=0.15$. A majority of the profiles have $t_a(\delta)=\textrm{08:00}$ irrespective of $\delta$ implying that a large proportion of occupants would have arrived in the buildings by 08:00 daily. On the other hand there is a variety of inferred $t_d(\delta)$ times as early as 16:00 and late as 23:00. Compared to fall profiles, summer profiles have earlier departure times. $t_d(\delta)$ is undefined for profile D-OF-3 when $\delta=0.05$ indicating that buildings B001, B002, B004 and B005 remain occupied beyond the last hour of the day on Monday -- Thursday in the fall when $\delta=0.05$. Similarly in building B003 whose fall Monday -- Thursday profile is D-OF-4, occupants remain after 23:00 on Monday -- Thursday when $\delta=0.05$ and $\delta=0.10$. We detect no occupancy in the Saturday -- Sunday summer occupancy schedule profile, D-OS-2 for all $\delta$ values. Increasing $\delta$ yields later and earlier $t_a$ and $t_d$ respectively resulting in narrower period of HVAC operation defined by $t_s^o(\delta,\tau)$ and $t_e^o(\delta,\tau)$.

\begin{table}
\centering
\caption{WiFi-derived first arrival time $t_a$, last departure time, $t_d$, HVAC ramp-up time, $t_s^o(\delta,\tau)$, setback time, $t_e^o(\delta,\tau)$ for $\tau=2$ and varied $\delta$.}
\label{tab:occ_time_signals}
\begin{tabular}{lrrrrrr}
\toprule
Profile &  $\delta$ &  $t_a$ &  $t_d$ &  $t_s^o(\delta,2)$ &  $t_e^o(\delta,2)$ \\
\midrule
 D-OS-1 &           0.05 &              08:00 &              19:00 &                06:00 &              17:00 \\
 D-OS-1 &           0.10 &              08:00 &              18:00 &                06:00 &              16:00 \\
 D-OS-1 &           0.15 &              09:00 &              17:00 &                07:00 &              15:00 \\
 D-OS-2 &           0.05 &             - &              - &               - &              - \\
 D-OS-2 &           0.10 &             - &              - &               - &              - \\
 D-OS-2 &           0.15 &             - &              - &               - &              - \\
 D-OS-3 &           0.05 &              08:00 &              20:00 &                06:00 &              18:00 \\
 D-OS-3 &           0.10 &              08:00 &              18:00 &                06:00 &              16:00 \\
 D-OS-3 &           0.15 &              09:00 &              18:00 &                07:00 &              16:00 \\
 D-OF-1 &           0.05 &              08:00 &              23:00 &                06:00 &              21:00 \\
 D-OF-1 &           0.10 &              08:00 &              21:00 &                06:00 &              19:00 \\
 D-OF-1 &           0.15 &              08:00 &              20:00 &                06:00 &              18:00 \\
 D-OF-2 &           0.05 &             10:00 &              22:00 &                08:00 &              20:00 \\
 D-OF-2 &           0.10 &             13:00 &              19:00 &               11:00 &              17:00 \\
 D-OF-2 &           0.15 &             15:00 &              16:00 &               13:00 &              14:00 \\
 D-OF-3 &           0.05 &              07:00 &              - &                05:00 &              - \\
 D-OF-3 &           0.10 &              08:00 &              22:00 &                06:00 &              20:00 \\
 D-OF-3 &           0.15 &              08:00 &              21:00 &                06:00 &              19:00 \\
 D-OF-4 &           0.05 &              08:00 &              - &                06:00 &              - \\
 D-OF-4 &           0.10 &              08:00 &              - &                06:00 &              - \\
 D-OF-4 &           0.15 &              09:00 &              23:00 &                07:00 &              21:00 \\
\bottomrule
\end{tabular}
\end{table}

\cref{tab:chilled_water_time_signals} summarizes $t_s^e$ and $t_e^e$ inferred from the chilled water demand profiles where all profiles but two have $t_s^e=05:00$. Also, there is homogeneity in detected $t_e^e$ times where it is 21:00 in all summer and fall profiles with the exception of D-CS-4 where it is 22:00. From visual inspection, 04:00 could be inferred to be $t_s^e$ for profiles D-CS-1 and D-CS-3 that could have been automatically captured by a hyper-parameter that defines a minimum change in demand to signify ramp-up. However, we find that while this approach may work well in detecting $t_s^e$, the descent in demand after the peak is gradual in most profiles hence, such hyper-parameter may not accurately detect $t_e^e$. Alternatively, $t_s^e$ and $t_e^e$ can be manually defined after visual inspection of the demand profiles. Nevertheless, our approach is conservative as it avoids minuscule changes in demand that will otherwise be detected and aims for the times of the most drastic changes to serve as $t_s^e$ and $t_e^e$.

\begin{table}
\centering
\caption{Chilled water demand profile-derived ramp-up time, $t_s^e$ and setback time, $t_e^e$.}
\label{tab:chilled_water_time_signals}
\begin{tabular}{lrr}
\toprule
Profile &  $t_s^e$ &  $t_e^e$ \\
\midrule
 D-CS-1 &                 05:00 &                       21:00 \\
 D-CS-2 &                         07:00 &                       21:00 \\
 D-CS-3 &                 05:00 &                       21:00 \\
 D-CS-4 &                         05:00 &                       22:00 \\
 D-CF-1 &         05:00 &                       21:00 \\
 D-CF-2 &         05:00 &                       21:00 \\
 D-CF-3 &                         08:00 &                       21:00 \\
 D-CF-4 &         05:00 &                       21:00 \\
\bottomrule
\end{tabular}
\end{table}

\begin{figure}
  \centering
  \includegraphics[width=\columnwidth]{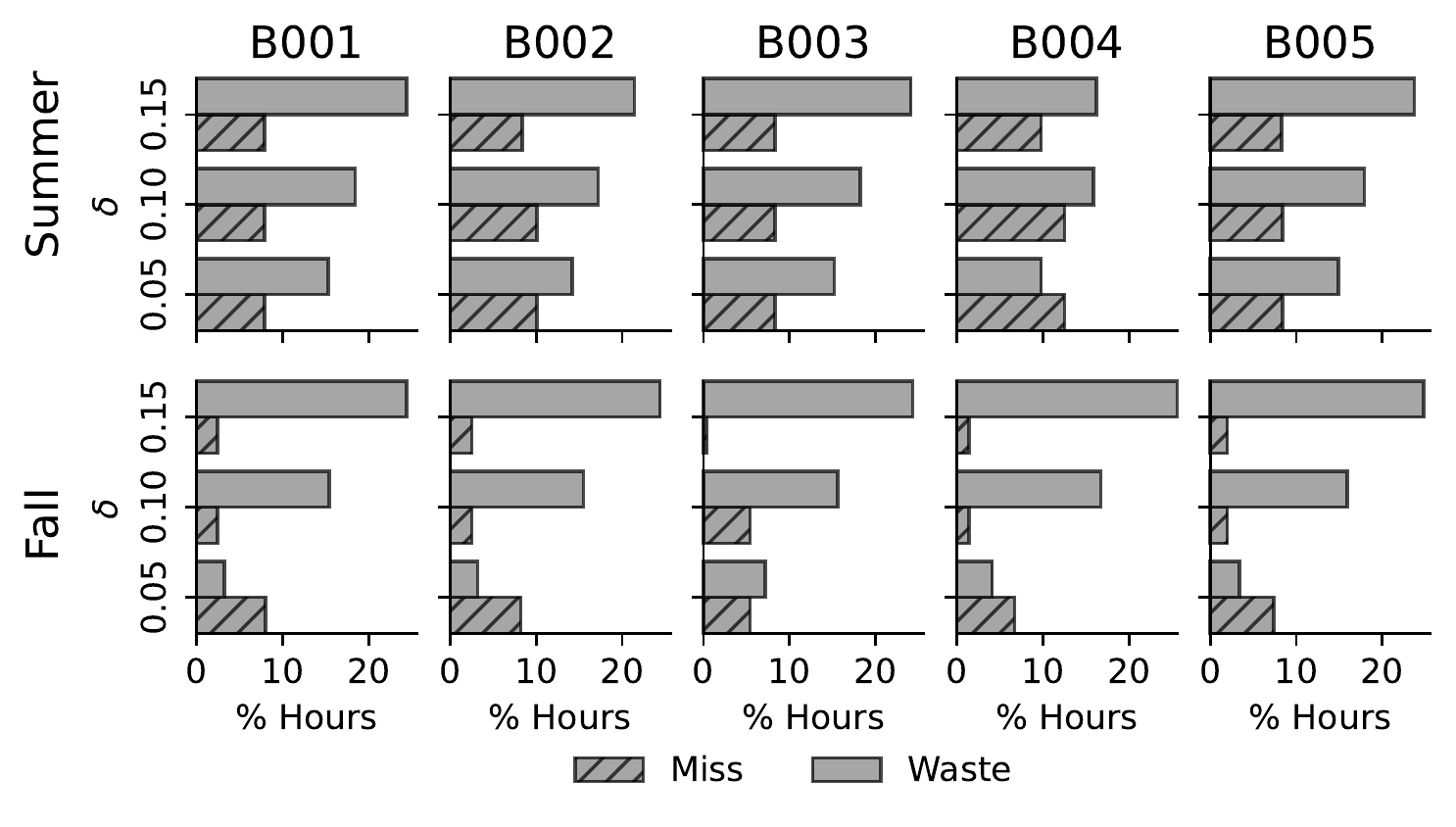}
  \caption{Miss and waste hours following~\cite{trivedi2017}.}
  \label{fig:miss_waste_hours}
\end{figure}

\citeauthor{trivedi2017} define waste and miss times as measures of energy savings and user comfort respectively when replacing static HVAC schedules with WiFi occupancy-derived HVAC schedules. \cref{fig:miss_waste_hours} summarizes these metrics when applied to our case study where waste hours occur when $t_s^o(\delta,\tau)>t_s^e$ or $t_e^o(\delta,\tau)<t_e^e$ and miss hours occur when $t_s^o(\delta,\tau)<t_s^e$ or $t_e^o(\delta,\tau)>t_e^e$. We see a directly proportional relationship between percentage of waste hours and $\delta$ in both semesters and all building where there is about 5\% and 10\% increase in waste hours per five percent increase in $\delta$ in the summer and fall semesters respectively. This relationship is as a result of narrower occupied periods at higher $\delta$ leading to shorter HVAC operation which saves energy that will otherwise be wasted if a static schedule were used. On the other hand, the percentage of miss hours remains almost unchanged at about 10\% irrespective of $\delta$ in all buildings during the summer. In the fall, miss hours are highest at $\delta=0.05$ and reduce or remain constant with increasing $\delta$ but, remain above zero. These observations align with our central hypothesis stated in \cref{sec:methodology} that if the building is operated within $t_s^o(\delta,\tau)$ and $t_e^o(\delta,\tau)$, unnecessary energy use is avoided without impacting occupant comfort.

\subsection{Energy Savings Estimation} \label{subsec:case_study-energy_savings_estimation}
Using MARTINI, we estimate chilled water energy savings from aligning demand with occupancy in the five D-OS and D-OF data sets' buildings and report our findings in \cref{tab:savings_with_occ}. We estimate an average of 8.1\% (64.9 MWh), 9.1\% (73.3 MWh) and 10.8\% (86.7 MWh) savings in the summer and 0.2\% (2.59 MWh), 3.3\% (63.0 MWh) and 5.9\% (109.0 MWh) savings in the winter when $delta=0.05$, $delta=0.10$ and $delta=0.15$ respectively. The average MWh savings in summer and fall are comparable except when $\delta=0.05$ and the reason for this is the extended nighttime occupancy on Monday -- Friday in the buildings at said $\delta$. Summer percent savings are larger owing to the fewer days in the summer compared to fall. In all cases but one, the buildings see reduced chilled water energy demand from aligning the HVAC operation with occupancy irrespective of $\delta$. B001, B003 and B005 have very similar savings at all $\delta$ values. There is increased demand of 0.1\% in B002 during the fall when $\delta=0.05$ while the highest percent savings from a single building occurs in B005 at 17.7\% in the summer when $\delta=0.15$. On average, We find increments of 1.3\% and 2.8\% in summer and fall average savings respectively for every 5\% increase in $\delta$.


\begin{table}
\centering
\caption{Estimated chilled water energy savings from aligning $t_s^e$ and $t_e^e$ with $t_s^o(\delta,\tau)$ and $t_e^o(\delta,\tau)$ for $\tau=2$ and varied $\delta$.}
\label{tab:savings_with_occ}
\begin{tabular}{lrrr}
\toprule
 & \multicolumn{3}{c}{$\delta$ (Summer)}\\
Building &     0.05      &  0.10       &    0.15       \\
\midrule
B001        &  11.8\% &  13.2\% &  15.8\%  \\
B002        &   3.1\% &   3.6\% &   4.2\%  \\
B003        &  10.8\% &  12.0\% &  14.3\%  \\
B004        &   1.4\% &   1.8\% &   2.0\%  \\
B005        &  13.5\% &  15.0\% &  17.7\%  \\
\midrule
\textbf{Average (\%)} & \textbf{8.1\%} & \textbf{9.1\%} & \textbf{10.8\%}  \\
\textbf{Average (MWh)} & \textbf{64.9} & \textbf{73.3} & \textbf{86.7}  \\
\bottomrule
 & \multicolumn{3}{c}{$\delta$ (Fall)}\\
Building &     0.05      &  0.10       &    0.15 \\
\midrule
B001        &  0.0\% &  3.3\% &  5.6\%  \\
B002        &  -0.1\% &  3.4\% &  5.9\%  \\
B003        &   0.9\% &  3.3\% &  6.6\%  \\
B004        &   0.1\% &  2.7\% &  4.6\%  \\
B005        &   0.1\% &  3.9\% &  6.7\%  \\
\midrule
\textbf{Average (\%)} & \textbf{0.2\%} & \textbf{3.3\%} & \textbf{5.9\%}  \\
\textbf{Average (MWh)} & \textbf{2.59} & \textbf{63.0} & \textbf{109.0}  \\
\bottomrule
\end{tabular}
\end{table}

\begin{figure*}
  \centering
  \includegraphics[width=\textwidth]{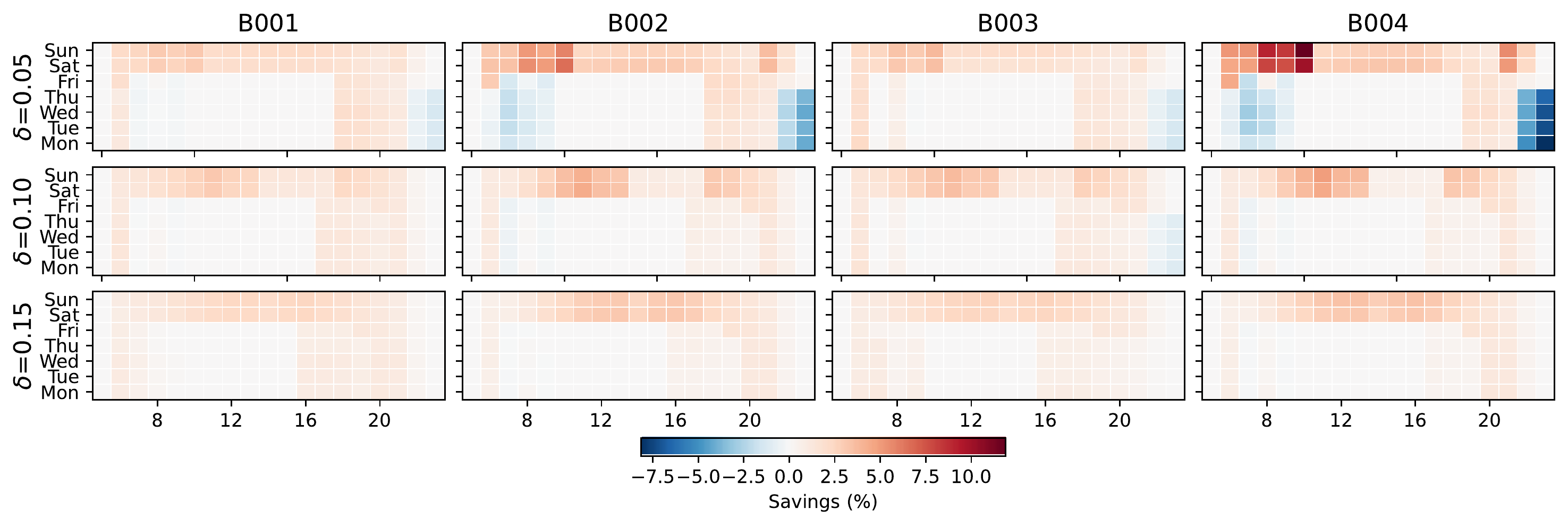}
  \caption{Hourly chilled water energy savings distribution.}
  \label{fig:hourly_savings_distribution}
\end{figure*}

\cref{fig:hourly_savings_distribution} shows how the savings from \cref{tab:savings_with_occ} are distributed throughout the week (combining summer and fall data). The saving distributions for B005 is similar to B001 hence, only the latter is shown. We see that majority savings are obtained during the weekends as a result of no/low occupancy during summer weekends and, late and early $t_s^o(\delta,2)$ and $t_e^o(\delta,2)$ respectively during fall weekends (see profiles D-OS-2 and D-OF-2 in \cref{tab:occ_time_signals}). Also, we observe that more savings are accrued during the evening setback start period. At $\delta=0.05$, there is increased demand on Monday -- Friday mornings owing to earlier $t_s^o(\delta,2)$ compared to $t_s^e$ for days when the chilled water energy demand is represented by profiles D-CS-2 or D-CF-3 and, nights as a result of extended occupancy detected in profiles D-OF-3 and D-OF-4. 

\begin{figure*}
  \centering
  \includegraphics[width=\textwidth]{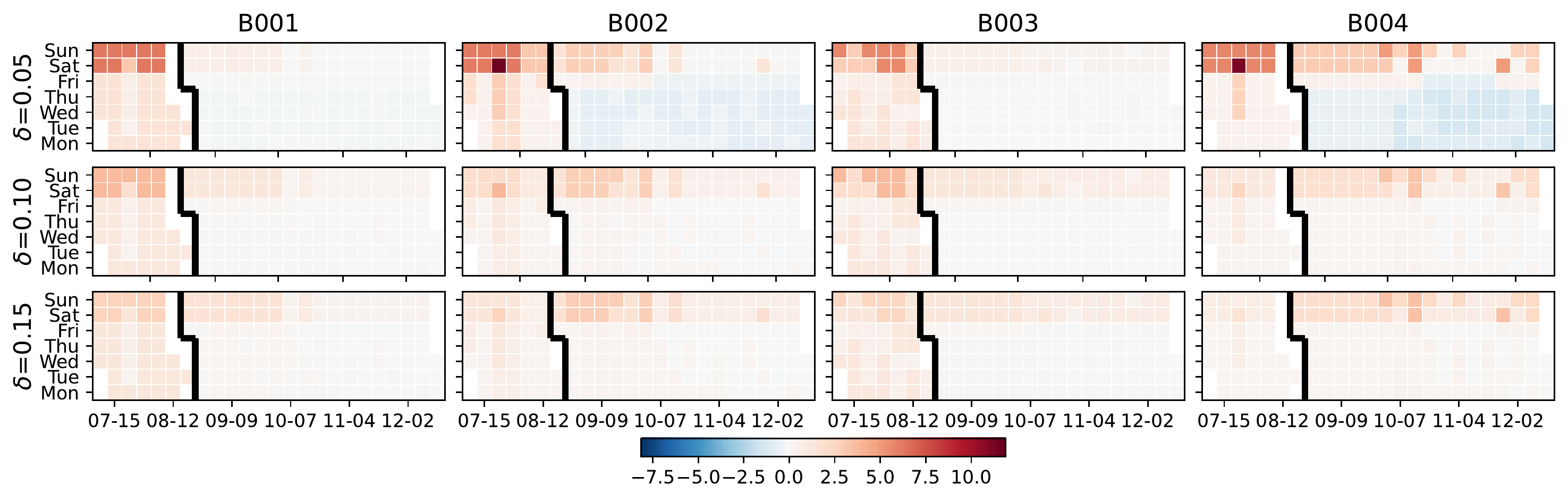}
  \caption{Daily chilled water energy savings distribution. Black line demarcates summer from fall.}
  \label{fig:daily_savings_distribution}
\end{figure*}

\cref{fig:daily_savings_distribution} shows the seasonality of energy savings in buildings B001, B002 and B004. Again B005 exhibits similar trend as B001 and is omitted. More savings come from the summer compared to the fall, most especially on weekends. At $\delta=0.05$, the fall period weekdays see increased demand instead of savings which, we attribute to weekday nighttime occupancy. At higher $\delta$, minuscule savings are derived from fall weekdays.

\subsubsection{BEPS Validation}
\begin{figure}
  \centering
  \includegraphics[width=\columnwidth]{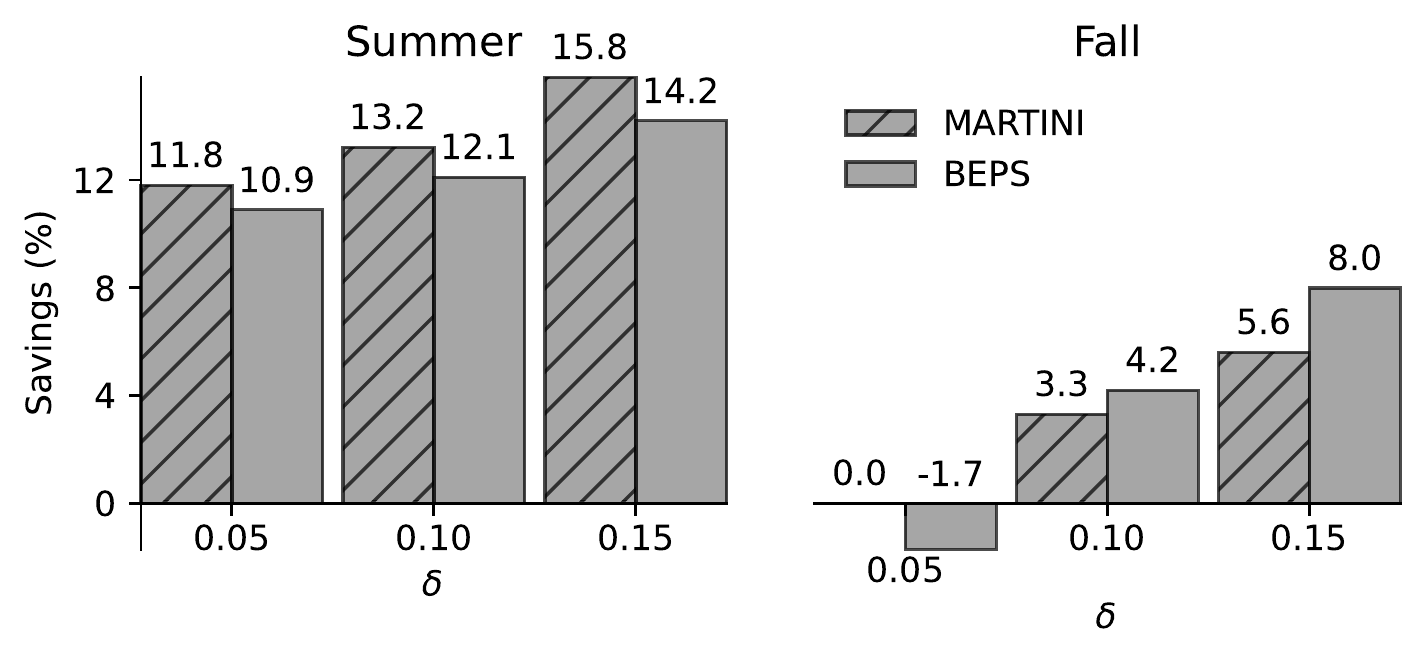}
  \caption{Comparison between MARTINI and building energy performance simulation chilled water energy savings estimation for B001.}
  \label{fig:beps_comparison}
\end{figure}

To validate our methodology, MARTINI, we compare the estimated chilled water energy savings in building B001 to that from a building energy performance simulation using WinAM \cite{esl_cc}. \cref{tab:beps_summary} provides a summary of B001's energy model where the building is served by three dual-duct variable air volume (DD-VAV) AHUs. The building's gross conditioned floor area is 16710 $m^2$. During unoccupied periods, the space cooling setpoint temperature is setup from 23.9 $^\circ C$ to 26.7 $^\circ C$ while the space heating setpoint temperature is setback from 21.1 $^\circ C$ to 18.3 $^\circ C$ in all AHU zones. In the baseline model, the control schedule is defined by $t_s^e$ and $t_e^e$ from B001's representative chilled water energy demand profiles D-CS-1, D-CS-3, D-CF-3 and D-CF-4. In the optimized models, $t_s^o(\delta,\tau)$ and $t_e^o(\delta,\tau)$ for varied $\delta$ with respect to B001's WiFi occupant-derived schedule in \cref{tab:building_occupancy_schedules} are used to re-define the control schedule. The relationship between the baseline model's chilled water energy demand and smart meter readings shows correlation coefficient (R\textsuperscript{2}) and Root Mean Squared Error (RMSE) of 0.93 and 248kWh, respectively, which we deem satisfactory. \cref{fig:beps_comparison} validates our approach showing that the savings estimated by MARTINI are within 0.9-1.6\% and 0.9-2.4\% of the BEPS savings projections in the summer and fall respectively. At $\delta=0.05$, BEPS estimates increased chilled water energy demand while it remains unchanged in MARTINI. It also shows that MARTINI tends to be somewhat conservative in the fall, and underestimates the savings, especially with increasing $\delta$. Lastly, we compare unmet hours per day during occupied periods in the optimized BEPS models to the baseline model and find about 50.0\% decrease in unmet hours/day during the summer and 36.3\% increase - 73.4\% decrease in unmet hours/day during the fall depending on what $\delta$ value is used. Ultimately $\delta$ is a hyper-parameter whose optimal value should be decided by a facility manager to trade-off between energy savings and occupant comfort.

\begin{table}
\centering
\caption{Building B001 WinAM energy model summary.}
\label{tab:beps_summary}
\begin{tabular}{lrrr}
\toprule
Parameter & AHU-1 & AHU-2 & AHU-3 \\
\midrule
System Type & DD-VAV & DD-VAV & DD-VAV \\
Floor Area ($m^2$) & 5890 & 4710 & 6110 \\
Wall U-Value ($\frac{W}{m^2K}$) & 0.857 & 0.857 & 0.857 \\
Window U-Value ($\frac{W}{m^2K}$) & 4.26 & 4.26 & 4.26 \\
Window-wall-ratio (\%) & 28.0 & 28.0 & 2.00 \\
Occupancy Peak (\#) & 358 & 166 & 364 \\
Design flowrate ($\frac{L}{s}$) & 37000 & 45700 & 40300 \\
\bottomrule
\end{tabular}
\end{table}

\subsubsection{Extrapolation \& Sensitivity Analysis}
\begin{figure}
  \centering
  \includegraphics[width=\columnwidth]{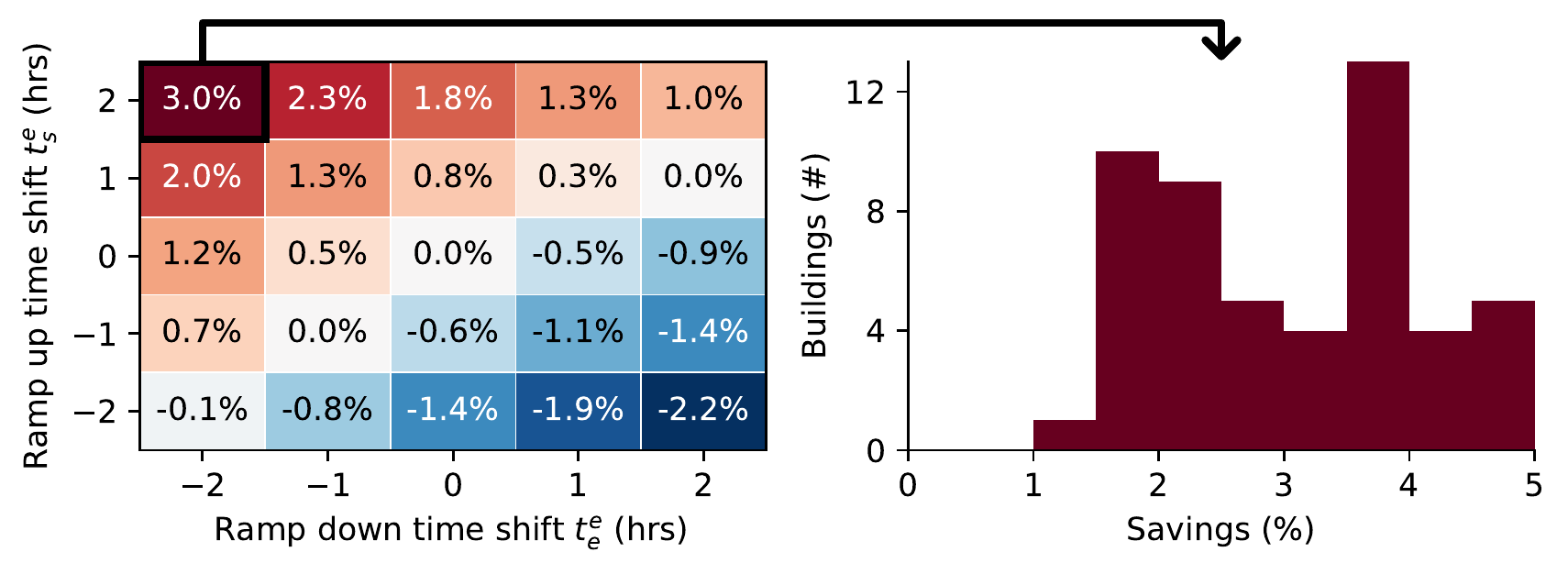}
  \caption{Average energy savings from shifting $t_s^e$ and $t_e^e$. The histogram shows distribution of the maximum average savings.}
  \label{fig:savings_from_shifting_without_occupancy_information}
\end{figure}

In \cref{fig:savings_from_shifting_without_occupancy_information}, we investigate the sensitivity of energy consumption to increasing or decreasing $t_s^e$ and $t_e^e$ in all 51 buildings. As expected, if we increase $t_s^e$ and decrease $t_e^e$, higher energy savings are accrued due to narrower period of HVAC operation. We find that an average of 3.0\% savings in chilled water energy demand is achievable if we shift $t_s^e$ forward and $t_s^e$ backward by two hours. From the distribution of the maximum savings at $t_s^e+2$ and $t_s^e-2$, we see that in about 17\% of the buildings, over 4\% savings is achieved, approximately 60\% of the buildings have 2\% - 4\% savings and the rest have at least 1\% savings potential.

\section{Discussion} \label{sec:discussion}
The University of Texas at Austin (UT) Sustainability Master Plan (SMP), 2016 outlines offsetting the campus space growth and related energy plant load growth as one of it's goals in the energy sub-area \citep{UtSmp16}. It puts forward, the implementation of new demand side strategic plan for energy conservation projects in existing buildings as a strategy towards achieving this goal and aimed to record 20\% reduction in energy use per square foot by the year 2020 with respect to a 2009 baseline \citep{UtSmp16}. The institution forecasts that a 2\% annual reduction in Energy Use Intensity (EUI) through to the year 2035 will keep peak demand below 70MW beyond the 2035 fiscal year, which is critical to offset the need for a costly power plant expansion \citep{DsspUem17}. Chilled water utility used to meet cooling demand on the UT campus, has the highest natural gas fuel cost and consumption per gross square meter at \$5.60/m\textsuperscript{2} and 0.365 Wh/m\textsuperscript{2} compared to electricity and steam utilities \citep{DsspUem17} hence is critical in attaining the annual EUI reduction target. While the UT's Utilities and Energy Management utilizes nighttime setback and setup ECM to obtain energy savings, it is yet to take advantage of available occupancy information offered by the institution's robust WiFi infrastructure in its application. MARTINI is designed to provide a tool for facility managers to generate value from their building's data and achieve institutional energy reduction goals.


We investigate the sensitivity of energy demand to shifted HVAC ramp-up and setback times in the absence of occupancy information for 51 buildings on the UT campus. We find that by increasing and decreasing the ramp-up and setback schedule start time by two hours, an average of 3.0\% chilled water energy savings is estimated where in 17\% of the buildings, over 4\% savings is achieved, approximately 60\% of the buildings have 2\%--4\% savings and the rest have at least 1\%. We recognize that while the energy savings are conservative, it's impact on the case-study campus' operational costs are significant. At an average utility rate of \$0.055/kWh, we estimate \$170,000.00 chilled water energy utility bill savings from just the 51 buildings out of the larger 178+ buildings on the campus. It is also, noteworthy to emphasize that these savings are achieved without having the need to install additional sensor infrastructure to capture occupant information, and our method provides an avenue to quickly arrive at an estimate on energy savings from occupant-driven nighttime setback scheduling without additional modeling or experimentation costs by leveraging available smart meter energy data.

A shortcoming of our methodology is that it may fail to captures the interaction between changing occupancy and other building heating and electricity utilities that may affect chilled water energy demand and overall building energy demand. For example, a shorter period of occupancy could mean reduced heat generating plug load that will directly reduce cooling demand but increase heating demand. A solution to this could be to develop a supervised regression model that predicts the other utilities using chilled water energy demand.

Other challenges to be addressed in our proposed framework include it's dynamic adaptation to changing building occupant pattern, building function and building energy management practices that will affect learned occupancy schedules and energy demand profiles. \citeauthor{trivedi2017} proposed two methods to solve these problems, specifically for HVAC scheduling using WiFi-derived occupancy data, namely; continuous or on-demand adaptation. In continuous adaptation, the schedules are recomputed each time new occupancy data is ingested while on-demand adaptation seeks to find the variance of new occupancy data from previously trained data and only recomputes the schedules if the variance is significant. In our application, a challenge in both adaptation methods is the selection of KMeans number of clusters, $k$ to be determined a priori, hence there is the need for manual intervention from a facility manager during validation in the learning stage.

There is the possibility of overestimated occupant counts which is critical when determining $t_a(\delta)$ and $t_d(\delta)$ based on $\delta$ as overestimated occupant counts will result in much earlier $t_a(\delta)$ or later $t_d(\delta)$ and diminished energy savings. A remedy that has been used in literature is to calibrate device counts against ground truth occupancy \citep{hobson2020,simma2019,8255034} however, this approach assumes static ratio of device counts to actual occupancy counts. \citeauthor{park2019good} investigated the use of Capture-Recapture, a methodology adopted from the field of zoology, to infer the spatial and temporal variation in WiFi device count to actual occupancy count ratio as an alternative to a statically determined ratio. Thus, our future work will explore the adoption of this approach as a correctional measure for overestimated counts.
\section{Conclusion} \label{sec:conclusion}
We introduce MARTINI, a sMARt meTer drIveN estImation of occupant-derived HVAC schedules and energy savings that leverages the ubiquity of energy smart meters and WiFi infrastructure in commercial buildings. MARTINI learns building occupancy schedules from clustered WiFi-sensed occupancy counts. Then, in conjunction with energy demand profiles learned from smart meter data, it provides estimate for energy savings from aligning HVAC ramp-up and setback times with actually observed occupancy. 

In our case study of five university buildings with occupancy data, we discover 8.1\%--10.8\% (summer) and 0.2\%--5.9\% (fall) chilled water energy savings when HVAC system operation is aligned with occupancy for a period of seven months. In the absence of occupancy information, we can still estimate potential savings from increasing ramp-up time and decreasing setback start time:We find savings potentials between 1\%--5\% in 51 academic buildings. We validate our method with results from building energy performance simulation (BEPS) and find that estimated average savings of MARTINI are within 0.9\%--2.4\% of the BEPS predictions.

MARTINI is a data-driven method that allows to quickly identify savings potentials in buildings without the need for a detailed energy model, and can be readily applied to building HVAC operation in large building portfolios.

\printcredits

\bibliographystyle{model1-num-names}

\bibliography{references}

\end{document}